# Implications of IR continua for x-ray emission/reflection in AGN

Gary J. Ferland, Peter G. Martin, Peter A.M. van Hoof, & Joseph C. Weingartner

## Abstract

Observations of infrared emission from AGN show that grains exist over a broad range of distances from the central object, extending to the point where they are destroyed by sublimation. These ~$10^3$ K grains produce much of the 1 μm continuum. In this region closest to the central object there must be a gaseous component associated with the hot grains. This paper employs a state of the art grain model and shows that the gas must be very hot, with temperatures in the neighborhood of $10^6$ K. The dusty component has a covering factor of roughly 50% and so this region also reprocesses much of the total x-ray emission. Our explicit models of the IR through x-ray spectral energy distributions allow the x-ray component to be predicted from IR observations. We are creating a grid of such predictions and will make them available as an XSPEC add-in, allowing this spectral component to be included in quantitative modeling of AGN spectra.

## 1 IR continua in AGN

The central regions of Active Galactic Nuclei (AGN) are strong x-ray sources, and modern instrumentation offers the opportunity to obtain high resolution spectroscopy for the first time. These spectra reveal rich complexity, in part because there are likely to be several sources of emission and reflection within the spatially unresolved central source. Given this unknown geometry all spectroscopic clues must be used to deduce what lies within the central parsecs.

Infrared observations show that dusty material exists over a broad range of distances from the central object, extending down to the distance from the central object where the grains reach their sublimation temperature (Sanders et al. 1989). This dust is thought to be the molecular torus that provides obscuration in unified AGN schemes. The luminosity of the observed infrared continuum shows that the hot grains subtend a covering factor $\Omega/4\pi \sim 0.5$ as seen from the central object (Rowan-Robinson 1995).

What are the spectroscopic properties of the layer containing the hot grains? Will it contribute to the observed x-ray, UV, or optical emission? The central regions contain several emission/reflection sources – could the gas associated with the hot grains be one of them? Can the x-ray emission be quantified and predicted from the infrared?

This paper builds upon existing models of the infrared continuum, using full photoionization simulations and our greatly improved treatment of the grain-gas interactions and grain emission, to show that the associated gas must be quite hot and is likely to be a significant source of x-ray emission.

## 2 Geometry

Grains roughly equilibrate at the energy density temperature of the local radiation field. Grains can extend up to the point where the grains sublimate. For a canonical sublimation temperature of 1000 K this point is (Rowan-Robinson 1995)

$$r(1000 \text{ K}) \approx 1 \text{ pc } L_{46}^{1/2}.$$

Grains can survive at smaller radii, depending on size and composition. For the hard radiation field emitted by the central source, silicates and smaller particles tend to be hotter than larger particles, or graphite, thus favoring larger graphite particles at small distances.

The form of the continuum shows that the regions are optically thick in the mid infrared. The inner several hundred parsec region therefore contains enough dust (covering factor about a half) to repeatedly reprocess the infrared continuum. The hottest grains must lie closest to the primary continuum source, which is taken to be the unextinguished continuum from the central object. Although the thermal reemission from the hot dust is intense, it occurs at longer wavelengths (1 – 10 μm). While it is important in heating more distant cooler grains, it has little effect on the ionization or temperature of the gas near the illuminated face of the torus.

This paper concentrates on the initial emission from the ionized layer of gas closest to the continuum source. In most unified schemes at least some lines of sight to the torus are relatively unobscured, so this emission component should be directly observed in some objects. We will show that it is likely to contribute to the observed x-ray emission.

The geometry we assume is based on that observed in the Orion star forming region (Baldwin et al. 1991; Ferland 2001). The H II region is a photoevaporative flow on the face of the background molecular cloud. The geometry can roughly be described as a hydrostatic layer with radiation and gas pressure held constant with increasing depth. The simple constant-pressure model reproduces the ultraviolet to infrared spectrum of Orion fairly well and is the one we use here.

## 3 Spectral simulations including hot grains

The calculations performed here are done with the current version of the spectral simulation code Cloudy, last described by Ferland et al. (1998). The grain model has been updated as described by van Hoof et al. (2001). In particular we explicitly include the grain physics described by Weingartner & Draine (2001) for interactions with the local radiation field and gas which heat, cool, and ionize the grains. The formalism also includes non-equilibrium emission (temperature spiking) for very small grains and can resolve the full grain size distribution. Additionally, a molecular network incorporating $H_2$, CO, and an additional 23 species allows the simulation to continue into predominantly molecular environments.

## 4 Calculations

### 4.1 Model parameters

We assume that the primary continuum is a simple power law with $f_\nu \propto \nu^{-1}$, extending from 10 μm to 50 keV. A total luminosity of $10^{46}$ erg s$^{-1}$ is taken and the inner radius is ultimately adjusted to 3 parsecs from the requirement that the calculated grain temperatures be no higher than ~1500 K. The density at the cloud's illuminated face is a free parameter. However, for a substantial fraction of the quasar's luminosity to be absorbed by *hot* grains, the primary continuum has to be absorbed over a distance comparable to the inner radius. Therefore, the density must be fairly high, $n > 10^4$ cm$^{-3}$. Abundances, dust to gas ratios, and silicate and graphite size distributions are taken as appropriate for the local ISM.

## 4.2 Conditions at the illuminated face

Figure 1 shows the computed gas and grain temperatures at the inner face for various values of the gas density. The temperatures were varied by changing the separation between the central source and the cloud. Clearly the gas associated with the warm grains is quite hot. Furthermore, the presence of the grains is important in establishing this temperature: a cloud with the same gas-phase abundances but without grains (or with the thermal effects of the grains suppressed) has a gas temperature that is roughly a factor of two lower. This is because the grains are a dominant opacity source, while the photoelectric opacity of the gas is diminished by its high ionization.

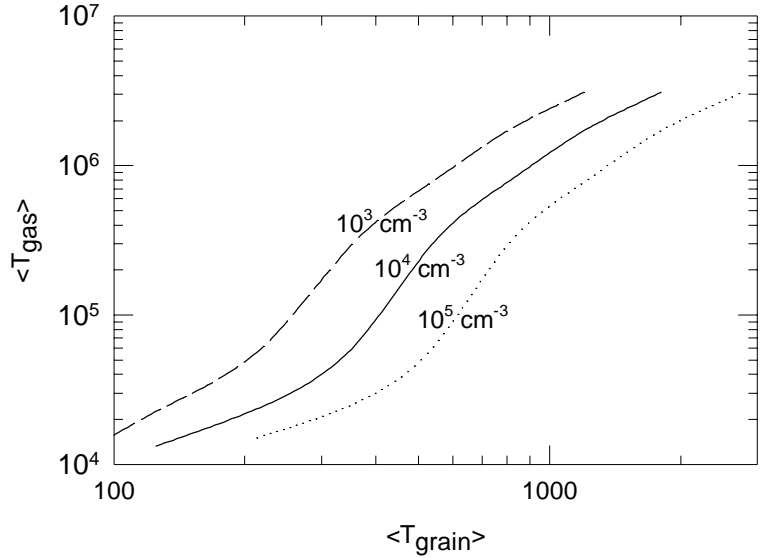

Figure 1. Loci of gas and grain (graphite component) temperatures at the illuminated face of the cloud, for various values of the gas density. Calculations were extended to temperatures above the grain sublimation temperature for illustration – these grains would have a short lifetime.

Figure 2 shows the computed gas temperature for various gas densities, with the inner radius adjusted such that the graphitic component has a temperature of $10^3$ K. The actual energy density temperature is ~600 K here, but the grains equilibrate at a somewhat greater temperature due to the fact that they are imperfect emitters. The temperature tends to decrease with increasing density since denser gas cools more efficiently.

Figure 3 shows the computed thermal structure of a cloud with a density at the illuminated face of $10^5$ cm$^{-3}$ and an ionization parameter of $10^2$ (this is for illustration only – many other choices of parameters are possible). Three distinct gas phases occur – a hot phase near the face of the cloud ($T \sim 10^6$ K), a "nebular" phase ($T \sim 10^4$ K) at middle distances, followed by a cold phase ($T < 10^3$ K) that is the shielded core of the cloud. The dust temperature also shows distinct steps (for these parameters the silicate component at the inner face is slightly above its sublimation temperature while the graphite is below it). Note that the relative extent of the nebular zone is the set by the increased density in this constant pressure model; the precise detail is not important to our discussion of the emission by hot grains and accompanying hot gas which arises in the relatively unobscured inner zone. Figure 4 shows the computed ionization and molecular structure involving oxygen, chosen as a representative constituent. It is successively fully stripped, atomic and singly ionized, and in molecular forms (mostly CO and $H_2O$) in the three thermal regions, respectively. Note that the radiation field is highly extinguished in the neutral zone (see below), and the combination of a relatively hard spectrum and the high density results in a warm but fairly neutral gas.

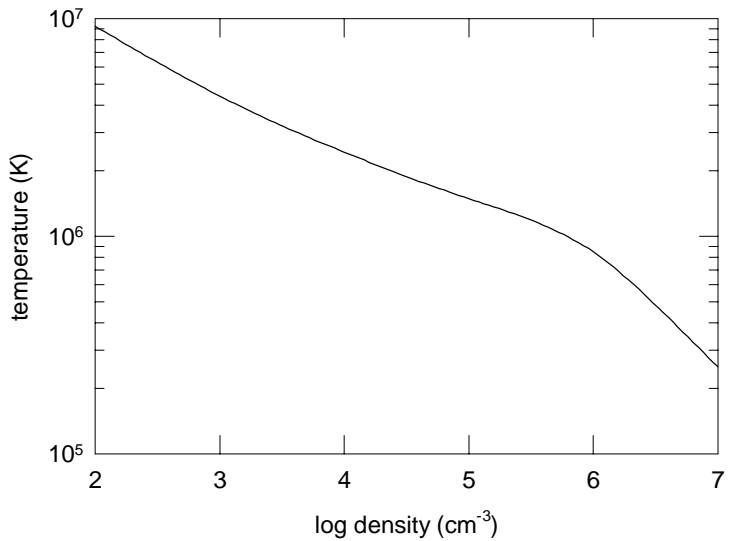

Figure 2. The gas temperature as a function of the gas density at the point in the model where the graphitic grain component has a temperature of $10^3$ K.

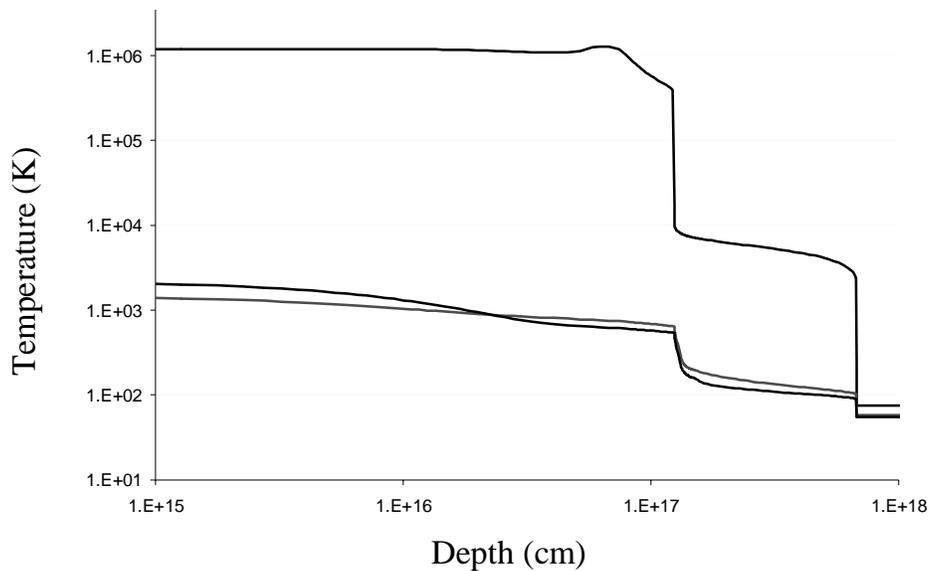

Figure 3 Thermal structure of full cloud. The thermal structure of the cloud with density $10^5$ cm$^{-3}$ is shown. The x-axis is the depth from the illuminated face. The upper curve gives the electron temperature and the lower red and black curves represent the silicate and graphitic components.

Grains and their associated opacity have a large effect on the shape of the radiation field and the structure of the cloud. This is a dust-bounded nebula – most ionizing radiation is absorbed by grains, not hydrogen. (The full optical depth of the computed structure is $A_V > 10^4$). This extinction will make the emission from the "nebular" region of the cloud (where T ~ $10^4$ K and most emission occurs in the optical) impossible to observe except at IR wavelengths.

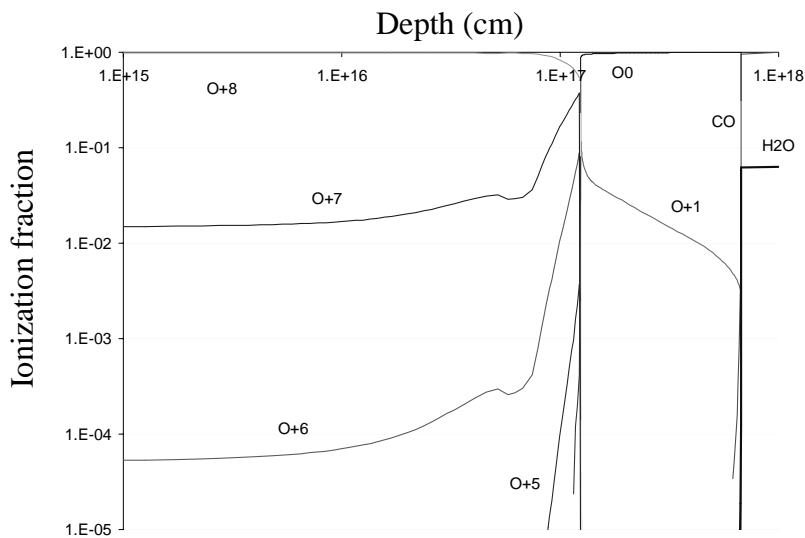

Figure 4 The computed ionization / molecular structure of oxygen. At the illuminated face O is fully stripped. It is predominantly atomic and singly ionized in the nebular regions, and locked in the form of CO and $H_2O$ at depth.

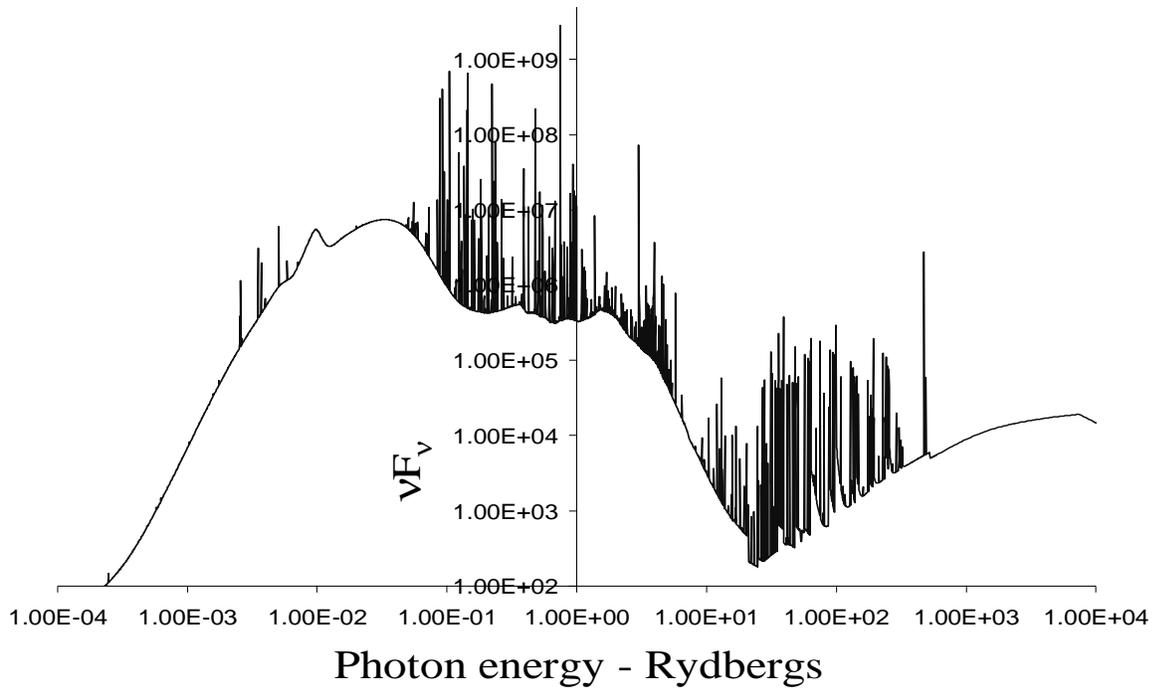

Figure 5. The spectrum emergent from the illuminated face of the model described here. The x-axis is the energy in Rydbergs and the y-axis is the flux per decade, $\nu F_\nu$, in surface brightness units (ergs cm$^{-2}$ s$^{-1}$).

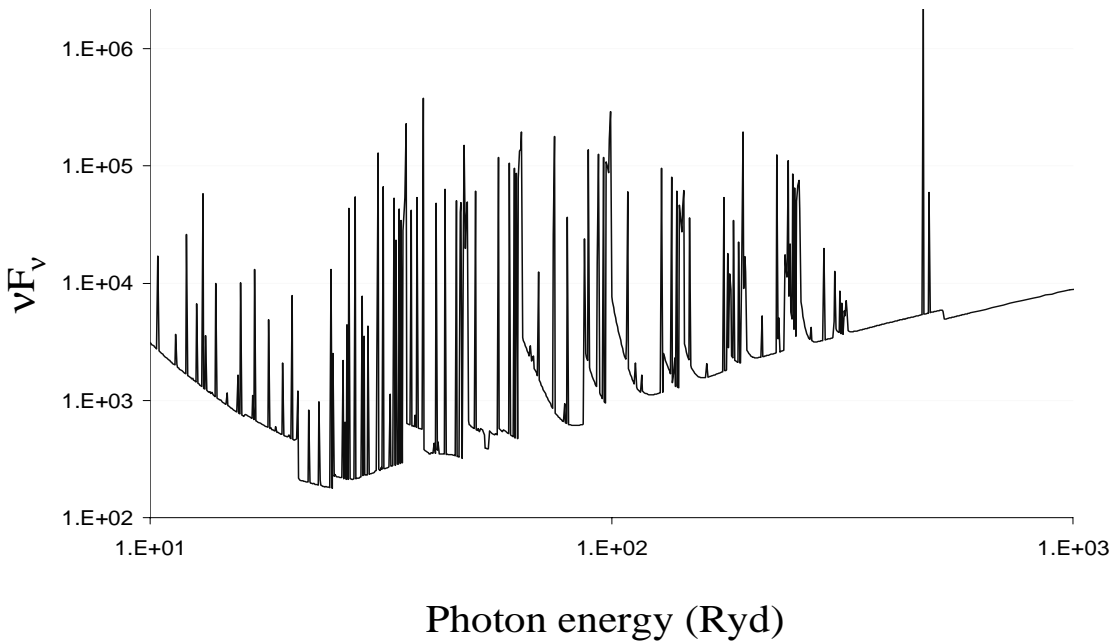

Figure 6  The x-ray spectrum that emerges from the computed structure. The lines are predominantly formed by recombination and a strong reflection component is present in the continuum.

Figure 5 and Figure 6 show the full spectrum that emerges from the inner face of our computed structure. Our goal is to use such calculations to make self-consistent predictions of the x-ray contribution from the molecular torus, given the observed hot grain near infrared emission.

# 5 Discussion and conclusions

It seems inescapable that the observed hot grains are accompanied by hot gas. We find gas temperatures typically ~ $10^5$ - $10^6$ K near the illuminated face. This hot plasma will contribute to the observed x-ray emission. (It is also possible that there is gas even closer to the central source, but the grains in it would have been sublimated, making it slightly cooler.) The nebular phase produces strong optical emission lines, but these form at depths with $A_V$ so large that this emission would be masked by other sources. At even greater depth in the cloud, the gas has become fully molecular, with $H_2$, CO, and $H_2O$ being major constituents. The molecules are warm, with T ~ 200 K, and should produce observable emission. The calculations show that the gas is Jeans unstable, with clumping on solar mass scales.

The predicted x-ray emitting gas has some interesting aspects. First, the temperature has in fact been enhanced by grain photoionization.

There is a bremsstrahlung continuum from the hot plasma, modified by the photoelectric absorption in the plasma through which it escapes. Furthermore, the electron scattering optical depth of the computed structure is in the range $\tau_e$ ~ 0.1 to 1. As a result a substantial part of the net emission from the emergent face is due to reflection (since the total covering factor of the torus must be ~1/2, reflection is significant in absolute terms too). The cloud also has a significant optical albedo, roughly 5%.

There is a rich x-ray line spectrum too. Fe provides an interesting example. In our model, Fe is highly depleted from the gas phase, in silicates. Any Fe in the gas phase would be highly stripped, with the Fe K$\alpha$ line occurring at ~6.7 keV. In this model the observed energy of the Fe K$\alpha$ line will correspond nearly to that of atomic Fe, 6.4 keV.

The detailed model shown here is only one possibility – parameters include the cloud density and its distance from the continuum source, with the constraint that hot grain emission be produced. We are in the process of producing extensive grids for correlating the x-ray and IR emission. The final product is an explicit prediction of the x-ray emission that must accompany the near infrared continuum. We will put these into XSPEC format so that a quantitative estimate of the contribution of the molecular torus to the observed x-rays can be made, by reference to the observed infrared.